\documentclass[acmsmall,screen]{acmart}

\acmJournal{PACMSE}
\acmVolume{1}
\acmNumber{1}
\acmArticle{1}
\acmYear{2026}
\acmMonth{1}



\AtBeginDocument{%
  \providecommand\BibTeX{{%
    \normalfont B\kern-0.5em{\scshape i\kern-0.25em b}\kern-0.8em\TeX}}}

\usepackage{microtype}
\usepackage{graphicx}
\usepackage{subfigure}
\usepackage{booktabs} 
\usepackage[ruled,vlined]{algorithm2e}

\usepackage{amsmath}
\usepackage{mathtools}
\usepackage{amsthm}
\usepackage{xspace}

\usepackage{multirow}
\usepackage{enumitem}
\usepackage{makecell}

\theoremstyle{plain}

\theoremstyle{definition}

\theoremstyle{remark}

\newcommand{\method}{\textsc{Prometheus}\xspace}

\usepackage{tcolorbox}
\tcbuselibrary{skins, breakable, theorems}
\usepackage{colortbl}
\usepackage{fancyvrb}
\usepackage{xcolor}
\usepackage{pifont}

\definecolor{darkgreen}{rgb}{0,0.5,0} 
\definecolor{purple}{RGB}{90,60,150} 
\definecolor{todocolor}{rgb}{0.9,0.1,0.1} 
\definecolor{fixcolor}{rgb}{0.1,0.7,0.3} 
\definecolor{hycolor}{rgb}{0.7,0.7,0.3} 
\definecolor{wycolor}{rgb}{0.9,0.1,0.1} 

\definecolor{lightred}{RGB}{255,225,220}
\definecolor{lightblue_node}{RGB}{52,119,203}
\definecolor{softyellow}{RGB}{200,140,20} 

\newcount\DraftStatus  
\newcommand{\nbc}[3]{\ifnum\DraftStatus=1
	{\colorbox{#3}{\bfseries\sffamily\scriptsize\textcolor{white}{#1}}}
	{\textcolor{#3}{\sf\small$\blacktriangleright$\emph{#2}$\blacktriangleleft$}}
\fi}

\newcommand{\draftnote}[2]{\ifnum\DraftStatus=1
	\marginpar{
		\tiny\raggedright
		\hbadness=10000
		\def\baselinestretch{0.8}
		\textcolor{#1}{\textsf{\hspace{0pt}#2}}}
\fi}

\newcommand{\parabf}[1]{\smallskip \noindent\textbf{#1.}\xspace}

\usepackage{babel}
\usepackage{hyperref}
\usepackage{caption}

\definecolor{lightcyan}{RGB}{10,110,150}
\hypersetup{
  colorlinks=true,
  linkcolor=lightcyan,
  citecolor=lightcyan,
  filecolor=lightcyan,
  urlcolor=lightcyan
}
\usepackage[capitalize,noabbrev]{cleveref}

\definecolor{lightblue}{RGB}{243,248,252}
\newenvironment{summary}{
\begin{tcolorbox}[width=\linewidth, colback=lightblue, top=1pt, bottom=1pt, left=2pt, right=2pt]

}
{
\end{tcolorbox}
}

\newtcbox{\capsule}[1]{%
  on line,
  colback=#1!15,
  colframe=#1!60!black,
  arc=5pt,
  boxrule=0.5pt,
  left=3pt,right=3pt,top=0.5pt,bottom=0.5pt
}

\title[Prometheus]{\method: Towards Long-Horizon Codebase Navigation \\ for Repository-Level Problem Solving}

\author{Yue Pan$^*$}
\affiliation{%
  \institution{University College London}
  \city{London}
  \country{United Kingdom}
}
\email{jack.pan.23@ucl.ac.uk}

\author{Zimin Chen$^*$}
\affiliation{%
  \institution{Sana Labs}
  \city{Stockholm}
  \country{Sweden}
}
\email{zimin@sanalabs.com}

\author{Siyu Lu}
\affiliation{%
  \institution{Uppsala University}
  \city{Uppsala}
  \country{Sweden}
}
\email{siyu.lu.6562@student.uu.se}

\author{Zhaoyang Chu}
\affiliation{%
  \institution{University College London}
  \city{London}
  \country{United Kingdom}
}
\email{zhaoyang.chu.25@ucl.ac.uk}

\author{Xiang Li}
\affiliation{%
  \institution{University College London}
  \city{London}
  \country{United Kingdom}
}
\email{x.li.25@ucl.ac.uk}

\author{Han Li}
\affiliation{%
  \institution{Nanjing University}
  \city{Nanjing}
  \country{China}
}
\email{231220161@smail.nju.edu.cn}

\author{Yang Feng}
\affiliation{%
  \institution{Nanjing University}
  \city{Nanjing}
  \country{China}
}
\email{Fengyang@nju.edu.cn}

\author{Claire Le Goues}
\affiliation{%
  \institution{Carnegie Mellon University}
  \city{Pittsburgh}
  \country{United States}
}
\email{clegoues@cs.cmu.edu}

\author{Federica Sarro}
\affiliation{%
  \institution{University College London}
  \city{London}
  \country{United Kingdom}
}
\email{f.sarro@ucl.ac.uk}

\author{Martin Monperrus}
\affiliation{%
  \institution{KTH Royal Institute of Technology}
  \city{Stockholm}
  \country{Sweden}
}
\email{monperrus@kth.se}

\author{He Ye$^{\dagger}$}
\affiliation{%
  \institution{University College London}
  \city{London}
  \country{United Kingdom}
}
\email{he.ye@ucl.ac.uk}

\thanks{$^*$Equal contribution. $^{\dagger}$Corresponding author. Email: he.ye@ucl.ac.uk.}

\begin{abstract}
Large Language Models (LLMs) have shown remarkable capabilities in automating software engineering tasks, spurring the emergence of coding agents that scaffold LLMs with external tools to resolve repository-level problems.
However, existing agents still struggle to navigate large-scale codebases, as the \textit{``Needle-in-a-Haystack''} problem persists even with million-token context windows, where relevant evidence is often overwhelmed by large volumes of irrelevant code and documentation.
Prior codebase navigation approaches, including embedding-based retrieval, file-system exploration, and graph-based retrieval, address parts of this challenge but fail to capture the temporal continuity of agent reasoning, rendering agents stateless and causing repeated repository traversals that hinder scalable planning and reasoning.

To address these limitations, we present \method, a memory-centric coding agent framework for long-horizon codebase navigation.
\method represents the repository as a unified knowledge graph to encode semantic dependencies and employs a context engine augmented with working memory that retains and reuses previously explored contexts to ensure continuity across reasoning steps.
Built upon this engine, \method integrates memory-enhanced navigation into a multi-agent system for automated issue resolution, encompassing issue classification, bug reproduction, patch generation, and verification.
Comprehensive experiments are conducted on two widely used issue resolution benchmarks, \textbf{i.e.}, SWE-bench Verified and SWE-PolyBench Verified.
Powered by GPT-5, \method achieves state-of-the-art performance with \textbf{74.4\%} and \textbf{33.8\%} resolution rates on the two benchmarks, ranking Top-6 and Top-1 among open-source agent systems, respectively.
Our data and code are available at \texttt{\url{https://github.com/EuniAI/Prometheus}}.
\end{abstract}

\begin{document}

\maketitle

\section{Introduction}

\textit{Large Language Models} (LLMs) have demonstrated strong capabilities in automating software engineering tasks, including code generation~\cite{liu2023llms_code_gen, jiang2023self_plan_code_gen, jiang2025code_generation_survey}, code summarization~\cite{sun2025code_summarization,virk2025code_summarization}, and program repair~\cite{zhang2024program_repair_review, ni2024next,rewardrepair,selfapr}.
Driven by this progress, coding agents have emerged as autonomous systems that scaffold LLMs with external tools to accomplish complex, repository-level tasks in real development environments~\cite{jimenez2024swebench, pan2025swegym}, as exemplified by SWE-agent~\cite{yang2024sweagent}, AutoCodeRover~\cite{zhang2024autocoderover}, and OpenHands~\cite{wang2025openhands}.

Despite substantial progress, existing coding agents still struggle to navigate large-scale codebases that comprise millions of lines of code and intricate dependency relationships. 
While recent LLMs have dramatically expanded their context windows to millions of tokens and beyond~\cite{openai2025gpt5, anthropic2025claudeopus45, deepmind2025gemini3pro}, these models still face the ``\textit{Needle-in-a-Haystack}'' challenge~\cite{liu2024lost_in_the_middle, hsieh2024ruler}. 
Specifically, task-relevant context evidence is overwhelmed by large volumes of irrelevant content, leading to inaccurate retrieval and degraded reasoning.
Here, context refers to code and accompanying documentation relevant to issue resolution.
As a result, effective codebase navigation for retrieving relevant context remains crucial for enabling long-horizon reasoning and decision-making throughout repository-level workflows.

\begin{figure*}[!t]
  \centering
  \includegraphics[width=0.98\textwidth]{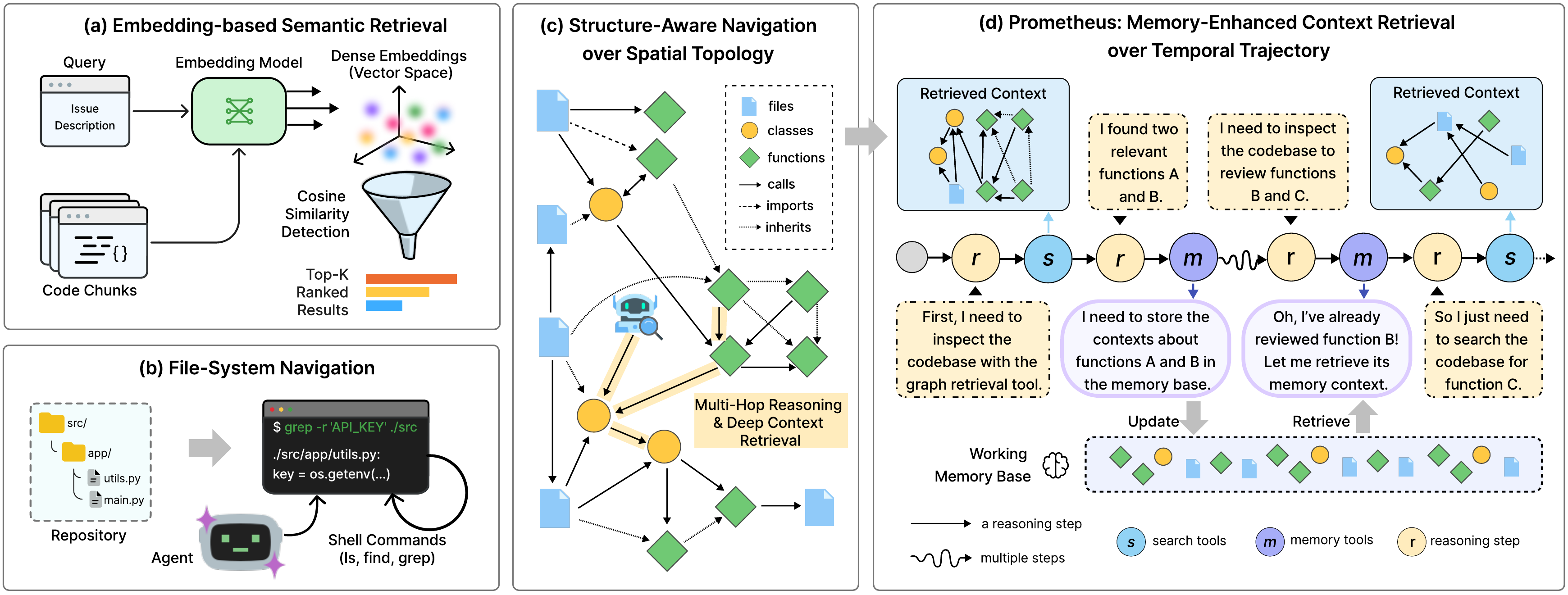}
  \caption{An illustration of existing codebase navigation methods for coding agents.}
  \label{fig_illustration}
\end{figure*}

\parabf{Existing Approaches and Limitations}
A straightforward navigation approach for coding agents is to apply embedding-based semantic retrieval, as illustrated in~\autoref{fig_illustration} (a), which encodes code chunks into embeddings and ranks them by their cosine similarity to the query embedding~\cite{xia2024agentless, zhang2023repocoder, reddy2025swerank, anysphere2026cursor}.
While enabling efficient navigation without step-by-step exploration, this approach relies heavily on chunking and embedding quality; irrelevant or noisy retrieval results can mislead the agent and degrade subsequent reasoning.
Another line of work adopts file-system navigation, as illustrated in~\autoref{fig_illustration} (b), which grounds context retrieval within the codebase by executing shell commands (\textit{e.g.}, \texttt{ls}, \texttt{find}, and \texttt{grep})~\cite{wang2025openhands, anthropic2025claudecode, anysphere2026cursor, openai2025codex} or interacting through specialized interfaces~\cite{yang2024sweagent, xia2025live}.
However, this method lacks a global view of the repository architecture, resulting in a fragmented understanding of the codebase that impairs long-horizon reasoning and planning. 
More importantly, both approaches are limited in handling the intricate semantic dependencies within the codebase, which are essential for repository-level problem solving.

Recently, structure-aware navigation has emerged as a promising alternative for coding agents~\cite{zhang2024autocoderover, tao2025code, ouyang2024repograph, yang2025kgcompass, chen2025locagent}, as shown in~\autoref{fig_illustration} (c), which models the repository as a graph, where nodes represent code entities (\textit{e.g.}, files, classes, and functions) and edges capture their dependencies (\textit{e.g.}, syntax trees, control flows, and data flows).
By explicitly encoding the hierarchical structure of the repository, these agents achieve powerful multi-hop reasoning and deep context retrieval across complex codebases.
While these agents effectively capture the spatial topology of the codebase, they fail to model the temporal history of their own exploration.
They operate essentially as stateless functions, lacking persistent memory of previously explored code.
As a result, agents repeatedly traverse the same regions of the repository, incurring substantial computational overhead.
Consequently, current agents remain inadequate for planning, reasoning, and memorizing required to navigate complex software systems on par with human developers.

\parabf{Our Work: Long-Horizon Codebase Navigation via Memory}
To address this gap, this paper introduces \method, a memory-centric coding agent framework that models codebase navigation from a temporal perspective, enabling long-horizon reasoning in repository-level workflows such as bug fixing, feature implementation, and refactoring.
Specifically, \method transforms the codebase into a unified knowledge graph that captures complex semantic dependencies and serves as the foundation for context retrieval.
Subsequently, \method employs a context engine augmented with working memory, as illustrated in~\autoref{fig_illustration} (d), which automatically retains and reuses previously explored contexts to maintain continuity across reasoning steps.
Finally, \method integrates the context engine into a multi-agent system for automated issue resolution, where specialized agents for issue classification, bug reproduction, patch generation, and patch verification operate sequentially under memory-enhanced navigation.

We conduct comprehensive experiments to validate the effectiveness of \method using the state-of-the-art LLM (\textit{i.e.}, GPT-5~\cite{openai2025gpt5}) on two widely used issue resolution benchmarks (\textit{i.e.}, SWE-bench Verified~\cite{jimenez2024swebench} and SWE-PolyBench Verified~\cite{rashid2025swepolybench}).
Experimental results demonstrate that \method achieves leading performance with 74.4\% and 33.8\% resolution rates on the two benchmarks, ranking Top-6 and Top-1 among open-sourced agent systems, respectively, at the time of writing.
In particular, \method achieves over 30\% balanced resolution rate across Java, Python, JavaScript, and TypeScript, indicating its strong multilingual generalization and consistent cross-language performance on SWE-PolyBench Verified.
Moreover, we conduct a fine-grained analysis of context retrieval performance by comparing retrieved contexts against human-annotated gold standards.
Results show that \method significantly outperforms leading agents, such as Agentless~\cite{xia2024agentless}, SWE-agent~\cite{yang2024sweagent}, and OpenHands~\cite{wang2025openhands}, retrieving a higher proportion of gold contexts at the file, function/class, and span levels under the same GPT-5 backbone.
Our ablation study further demonstrates that the working memory mechanism reduces LLM invocation cost by 45.4\% while improving the issue resolution rate by 25\%.

\parabf{Contributions}
The primary contributions of this paper are summarized as follows.

\begin{itemize}[leftmargin=4mm, itemsep=0.05mm]

\item 
\textbf{Memory-Enhanced Context Retrieval.}
We propose a novel context retrieval approach augmented with working memory, which models the temporal trajectory of agent navigation.
This mechanism enables agents to automatically retain and reuse previously explored contexts, achieving continuity and coherence in long-horizon reasoning.

\item
\textbf{Prometheus: A Novel Issue Resolution Agent.}
We develop \method, a multi-agent system that integrates memory-enhanced navigation into issue resolution workflows.
This system coordinates specialized agents for issue classification, bug reproduction, patch generation, and patch verification, establishing a unified workflow for automated issue resolution at the repository level.

\item
\textbf{Extensive Evaluation and Analysis.}
We conduct comprehensive experiments on two widely used issue resolution benchmarks (\textit{i.e.}, SWE-bench Verified~\cite{jimenez2024swebench} and SWE-PolyBench Verified~\cite{rashid2025swepolybench}), demonstrating the superiority of \method over existing agent systems.
We further demonstrate the superiority of \method in both context retrieval and multilingual adaptability, and validate the effectiveness of the working memory mechanism.

\end{itemize}

\section{Methodology}

\begin{figure*}[!t]
  \centering
  \includegraphics[width=0.98\textwidth]{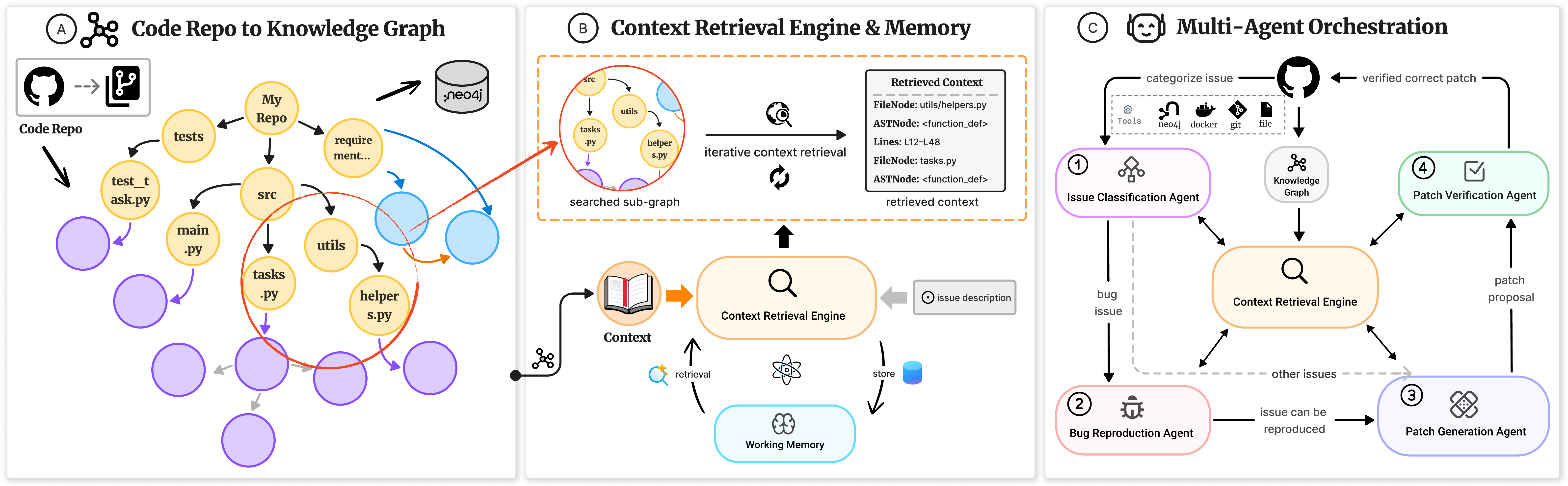}
  \caption{An overview of \method.}
  \label{method}
\end{figure*}

\autoref{method} presents an overview of \method, which is composed of three core components:
\textbf{(A) Repository to Knowledge Graph}, which constructs a unified code index representing the structural and semantic relationships within the target codebase;
\textbf{(B) Memory Enhanced Context Retrieval}, which enables effective identification and reuse of relevant code and documentation to support reasoning; and
\textbf{(C) Multi-Agent Architectural Design}, which coordinates specialized agents to collaboratively perform issue resolution tasks.
The following sections describe each component in detail.

\subsection{Repository to Knowledge Graph}
\label{graph-sec}

\begin{figure*}[!t]
  \centering
  \includegraphics[width=0.98\textwidth]{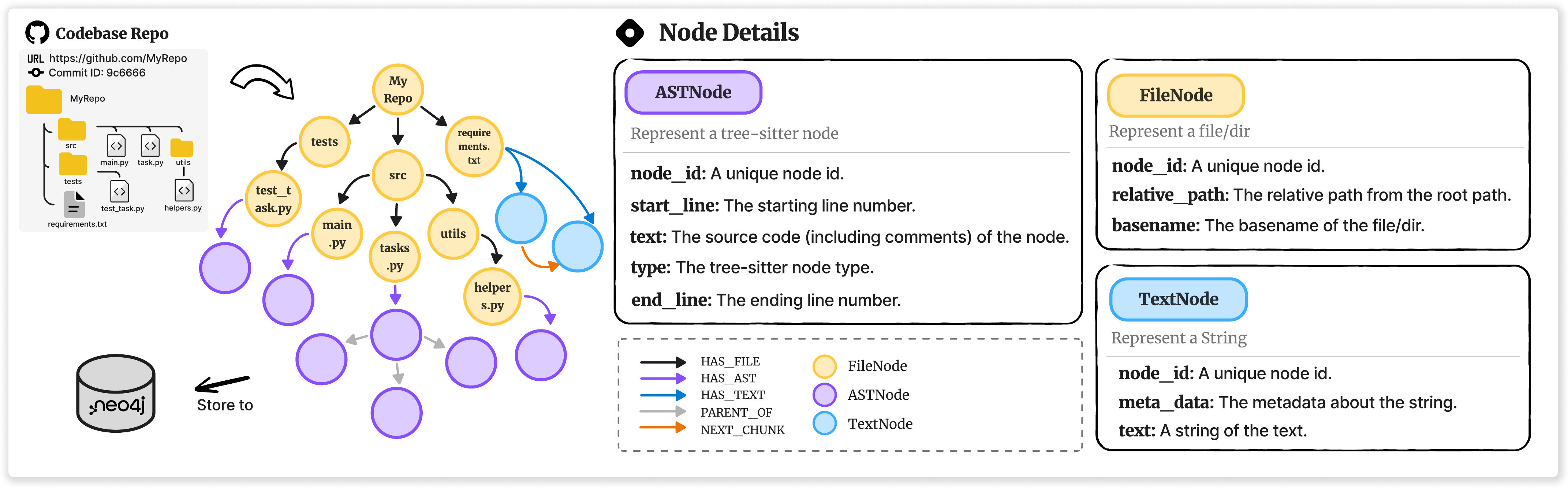}
  \caption{An overview of our knowledge graph construction for the codebase.}
  \label{knowledge_graph}
\end{figure*}

To facilitate semantic understanding and context retrieval across large-scale codebases, we propose a unified knowledge graph representation that integrates file structures, ASTs, and textual content into a coherent graph abstraction. As shown in~\autoref{knowledge_graph}, our knowledge graph is built around three core components: (1) defining a node and edge schema, (2) constructing the graph from source files, and (3) persisting the graph data in a scalable database.

\paragraph{Graph Schema}
The knowledge graph represents codebases as heterogeneous graphs composed of three primary node types: \ding{182} \capsule{softyellow}{FileNode}, \ding{183} \capsule{purple}{ASTNode}, and \ding{184} \capsule{lightblue_node}{TextNode}. 
A \texttt{FileNode} represents a file or directory with three attributes: a unique \texttt{node\_id}, the \texttt{relative\_path} from the repository root, and the \texttt{basename} of the file or directory. It anchors structural links in the knowledge graph.
An \texttt{ASTNode} represents a Tree-sitter syntax node. It includes a unique \texttt{node\_id}, the \texttt{start\_line} and \texttt{end\_line} indicating its position in the source file, the text of the code it covers (including comments), and its type, per the Tree-sitter grammar node type.
A \texttt{TextNode} represents a chunk of unstructured textual content in the knowledge graph. Each \texttt{TextNode} is associated with a unique \texttt{node\_id}, descriptive \texttt{meta\_data}, and the corresponding text span. TextNodes are extracted from documentation-oriented file types, including \texttt{.markdown}, \texttt{.md}, \texttt{.txt}, and \texttt{.rst}. The text is segmented into chunks based on a configurable \texttt{chunk\_size}. Depending on the file content and structure, adaptive splitters are applied using hierarchical separators, including paragraph-level (\texttt{\textbackslash n\textbackslash n}), line-level (\texttt{\textbackslash n}), word-level (space), and character-level splitting.

To capture relationships across FileNodes, ASTNodes, and TextNodes, we define five directed edge types.
These relationships are essential for representing the structural, syntactic, and lexical context necessary for code understanding and issue resolution. 
 The \texttt{HAS\_FILE} edge (black) connects directories to their child files or subdirectories, preserving the repository hierarchy. The \texttt{HAS\_AST} edge (purple) links each file node to the root of its corresponding abstract syntax tree. \texttt{PARENT\_OF} edges (gray) connect AST nodes to reflect syntactic hierarchy within the tree. \texttt{HAS\_TEXT} edges (blue) associate file nodes with their segmented textual content, and \texttt{NEXT\_CHUNK} edges (orange) connect sequential text chunks to maintain document order. 
 Together, these relationships enable the graph to represent structural, syntactic, and lexical information in an integrated and practical format, exposing the repository knowledge graph through a set of structured query tools.

To our knowledge, this work introduces a highly extensible knowledge graph schema
that unifies repository structure, program syntax, and unstructured text. In contrast to prior graph-based approaches~\cite{ouyang2024repograph} that rely on specialized or task-coupled designs, we adopt a deliberately minimal yet expressive schema, consisting of only three node types and a small set of well-defined relations. 
On SWE-bench Verified~\cite{jimenez2024swebench}, we observe an average knowledge graph construction time of 1.99 seconds per instance over all 500 instances, indicating that the proposed schema is lightweight enough for practical, large-scale use. This simplicity enables efficient graph construction and incremental updates as codebases evolve, while preserving sufficient structural and semantic coverage for repository-level reasoning. Consequently, the resulting knowledge graph provides a reusable substrate for tasks such as semantic retrieval, long-horizon dependency analysis, and agent-based code navigation, without requiring task-specific re-engineering.

\subsection{Memory-Enhanced Context Retrieval}

\subsubsection{Context Retrieval Engine}

\begin{figure*}[!t]
  \centering
  \includegraphics[width=0.98\textwidth]{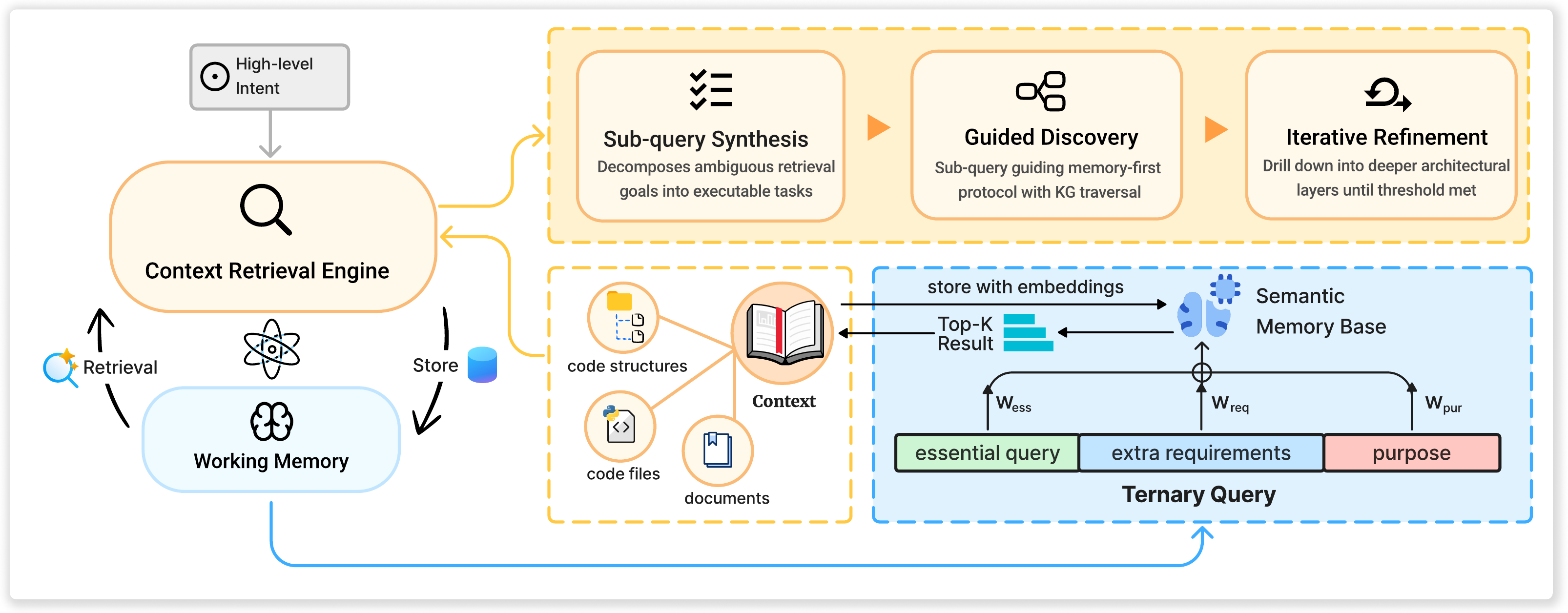}
  \caption{An overview of the memory-enhanced context retrieval engine.}
  \label{context_retrieval_engine}
\end{figure*}

\begin{figure*}[!t]
  \centering
  \includegraphics[width=0.98\textwidth]{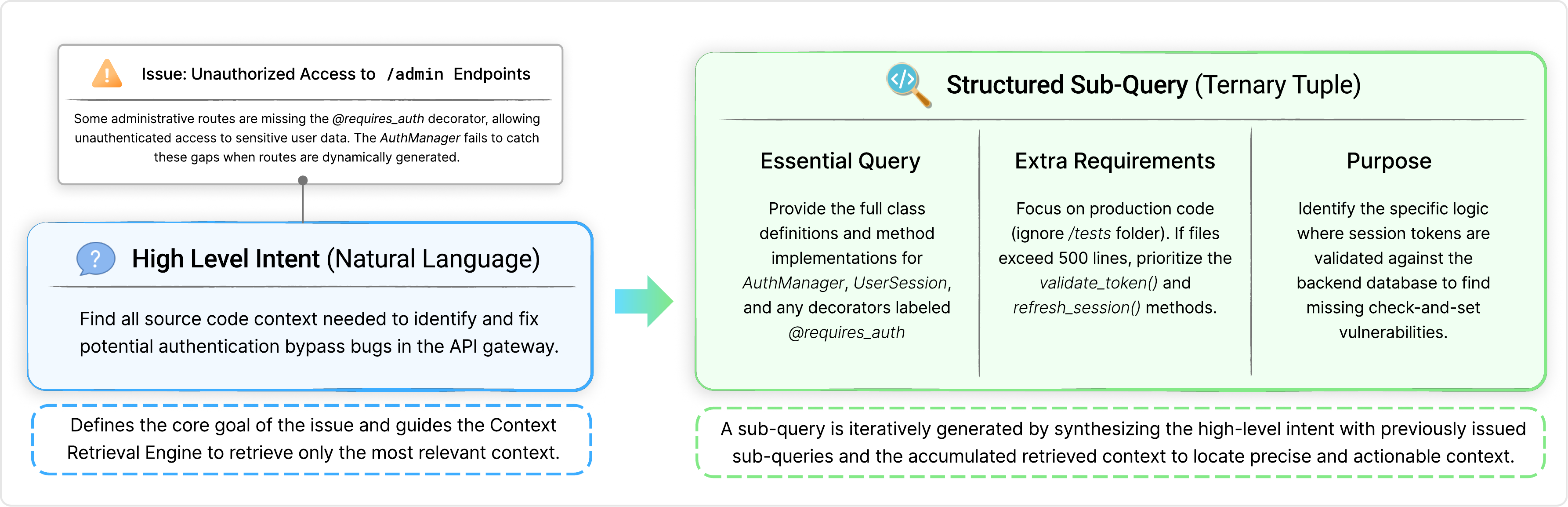}
  \caption{An example of high-level intent and sub-query.}
  \label{examples_high_level_intent_subquery}
\end{figure*}

As shown in~\autoref{context_retrieval_engine}, the Context Retrieval Engine adopts a three-stage retrieval cycle that incrementally maps a high-level intent expressed in natural language to structured contextual evidence. As summarized in Algorithm~\ref{alg:context-retrieval-engine}, the process begins with Sub-query Synthesis, where a structured and specific sub-query in natural language is generated with high-level intent and the current collected context. The sub-query is formulated as a ternary tuple comprising an \textit{essential query}, which captures the core information needed, \textit{extra requirements}, which encode retrieval constraints such as prioritizing specific files or truncating oversized artifacts, and a \textit{purpose}, which specifies the motivation or objective of the retrieval. \autoref{examples_high_level_intent_subquery} provides an example and illustration for both high-level intent and its corresponding sub-query. For example, a high-level intent to ``locate authentication bugs'' can generate a structured query targeting authentication-related classes, decorators, and their associated call sites.

Following synthesis, the context engine performs sub-query guided context retrieval and context organization under a memory-first protocol. As shown in Algorithm~\ref{alg:context-retrieval-engine}, the system first retrieves from the working memory for previously stored context associated with the current sub-query. If no relevant context is retrieved, the engine exposes the repository knowledge graph through a set of structured search tools, allowing the LLM to iteratively traverse file-level and AST-level nodes via multi-hop relations. Guided by the sub-query, the LLM autonomously decides which graph relations to follow and when to expand or stop traversal. For the set of discovered contexts, the engine first performs a structural deduplication by cross-referencing their file paths and line-number intervals ($start\_line, end\_line$). These contexts are categorized relative to each other as \textit{duplicate}, \textit{contained}, \textit{contains}, or \textit{separate}. For those identified as non-redundant, the LLM then performs an extraction step to distill the most relevant information. The engine keeps the code’s structure intact by grouping snippets by their source files and restoring their original order, forming a validated context. The validated contexts is then stored in memory for future use. This process is repeated and dynamically updated.

\subsubsection{Working Memory}
The goal of working memory is to prevent \method from repeated query and reduce the token cost.
As shown in~\autoref{context_retrieval_engine}, the working memory system in \method implements a structured and persistent context layer for repository-level evidence. We define a \emph{context unit} as a minimal reusable piece of context, represented as a tuple of (\textit{code snippet}, \textit{metadata}), where metadata includes the originating file path and start/end line numbers. Working memory maintains a set of \emph{memory records}, each functioning as a logical container that binds a structured \textit{sub-query} (as synthesized by the Context Engine) to its corresponding \textit{context units}. This design ensures that retrieval is not treated as a monolithic string search; instead, subsequent reasoning steps can perform fine-grained lookups by aligning their intent with the structured sub-queries preserved within the memory records. By the definition of agent memory in recent surveys on agentic memory systems~\cite{hu2025memory}, this module is classified as an agent memory component rather than a standard RAG module, because it maintains a persistent and continually updated internal memory instead of retrieving from a static external corpus.

\paragraph{Storage.}
Unlike conventional agentic memory that embeds a single concatenated query into one vector, \method adopts a Multi-vector Weighted Retrieval design. For each synthesized sub-query comprising three distinct fields—\textit{essential query}, \textit{extra requirements}, and \textit{purpose}—the storage pipeline computes semantic embeddings for these fields using a configurable embedding model API. This process yields a 3-vector representation for each memory record, which is then persisted in a database. Each vector is stored in a dedicated column, denoted as $\mathbf{e}_{ess}$, $\mathbf{e}_{req}$, and $\mathbf{e}_{pur}$, representing the \textit{essential query}, \textit{extra requirement}, and \textit{purpose} of the sub-query, respectively. These are stored alongside their corresponding text fields for traceability and debugging. These memory records are linked to their associated context units, and insertion is executed as a single database transaction (memory record, context unit, and linkage) to ensure referential integrity under concurrent agent execution.

\paragraph{Retrieval.}
During retrieval, the system encodes the structured input query into the same set of query vectors ($\mathbf{q}_{ess}, \mathbf{q}_{req}, \mathbf{q}_{pur}$). Prior work~\cite{khattab2020colbert} has shown that aggregating multiple similarity signals via weighted linear combination is a standard and theoretically well-founded practice in information retrieval. For a candidate memory record $m$, we compute a weighted similarity score by aggregating cosine similarities across the three vectors:
\[
s(m)=w_{ess}\cdot \cos(\mathbf{q}_{ess},\mathbf{e}^{m}_{ess})+
w_{req}\cdot \cos(\mathbf{q}_{req},\mathbf{e}^{m}_{req})+
w_{pur}\cdot \cos(\mathbf{q}_{pur},\mathbf{e}^{m}_{pur}),
\]
where $w_{ess}, w_{req}, w_{pur}$ control the relative importance of each intent component. To accelerate nearest-neighbor search, we build \textit{IVFFlat} indexes over the vector columns and perform retrieval in two stages: (i) run approximate nearest-neighbor (ANN) search independently for each vector column to obtain top-$n$ candidates per component; (ii) take the union of candidates and re-rank them using the aggregated score $s(m)$. Finally, we filter results below a similarity threshold and return the top-$k$ memory records, whose linked context units are provided as reusable evidence for downstream reasoning. This multi-vector weighted retrieval improves alignment between query intent and stored evidence, enhancing precision (by respecting auxiliary constraints) and recall (by capturing purpose-level signals) compared to single-vector agentic memory baselines.

\subsection{Multi-Agent Architectural Design}

\begin{algorithm}[t!]
\caption{Context Retrieval Engine with Single-Query Refinement}
\label{alg:context-retrieval-engine}
\KwIn{High-level retrieval intent $\mathcal{Q}$, Knowledge Graph $\mathcal{G}$, Working Memory $\mathcal{M}$, maximum query number $T_{\max}$}
\KwOut{Refined contextual evidence set $\mathcal{C}$}

Initialize empty context set $\mathcal{C}$\;
Initialize refinement step counter $t \gets 0$\;

\textcolor{green!40!black}{// Stage 1: Sub-query Synthesis}\;
Synthesize a single structured sub-query $q = (q_{ess}, q_{req}, q_{pur})$ from $\mathcal{Q}$\;
\textcolor{green!40!black}{// $q_{ess}$: Essential Query; $q_{req}$: Extra Requirements; $q_{pur}$: Purpose}\;

\While{q is not empty \textbf{and} $t < T_{\max}$}{
    \textcolor{green!40!black}{// Stage 2: Structural Discovery \& Context Organization (Memory-First)}\;
    Query context $\mathcal{C}_q$ from Working Memory $\mathcal{M}$ with sub-query $q$\;
    \If{not $\mathcal{C}_q$}{
        Traverse Knowledge Graph $\mathcal{G}$ guided by $q_{ess}$ and $q_{req}$\;
        Extract structural entities (e.g., FileNodes, ASTNodes)\;
        Group retrieved contexts by source file\;
        Sort contexts within each file by $(start\_line, end\_line)$\;
        Persist organized context $\mathcal{C}_q$ into $\mathcal{M}$\;
    }
    Merge $\mathcal{C}_q$ into global context set $\mathcal{C}$\;
    \textcolor{green!40!black}{// Stage 3: Iterative Context Refinement}\;

    Refine sub-query $q$ by updating $(q_{ess}, q_{req}, q_{pur})$\;
    \textcolor{green!40!black}{// e.g., drill down into caller hierarchies or dependency chains}\;

    $t \gets t + 1$\;
}

\Return{$\mathcal{C}$}
\end{algorithm}

\begin{figure*}[!t]
  \centering
  \includegraphics[width=0.98\textwidth]{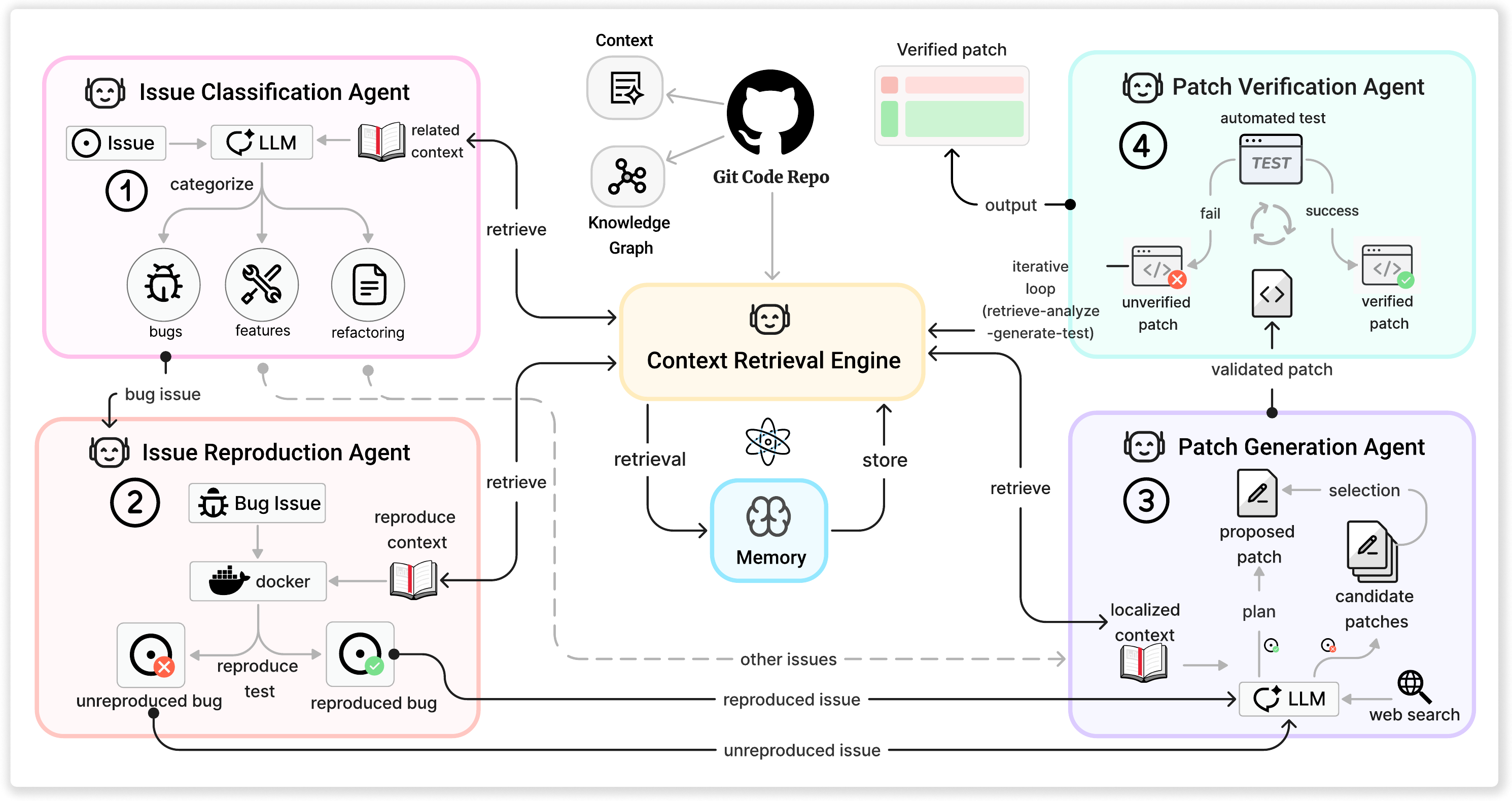}
  \caption{An overview of our multi-agent architecture.}
  \label{multi_agent}
\end{figure*}

As shown in~\autoref{multi_agent}, the architecture of \method consists of four primary functional agents, unified by a workflow. The orchestration follows a pipeline from issue classification to verified resolution, where the context and retrieved knowledge are shared across the Issue Classification Agent, Bug Reproduction Agent, Patch Generation Agent, and Patch Verification Agent.

\subsubsection{Issue Classification Agent}

The Issue Classification Agent serves as the intelligent gateway of \method, tasked with analyzing and categorizing incoming repository issues into distinct types, such as bug, feature, or refactoring. By performing this context-aware classification prior to repair planning, the agent enables the system to route heterogeneous issues to specialized downstream workflows, ensuring that each problem type is addressed by a tailored resolution strategy. This architectural choice constitutes a key novelty of our system. Unlike prior automatic program repair pipelines that are narrowly scoped to bug fixing, our approach explicitly generalizes issue understanding to reflect real-world software maintenance practices. By moving away from a uniform bug-fixing paradigm, \method can function as a general-purpose agentic issue resolution system, capable of handling the diverse mixture of problem types found in modern issue trackers.

\subsubsection{Bug Reproduction Agent}

The Bug Reproduction Agent aims to automatically construct a minimal yet executable reproduction of reported bugs, serving as a concrete and verifiable foundation for subsequent repair. Given an issue report consisting of the title, description, and discussion history, the agent first synthesizes a structured reproduction context that captures the expected failure behavior and relevant environmental assumptions. In real-world issue resolution, such a reproduction is indispensable, as it transforms ambiguous natural language descriptions into a deterministic test oracle. This design reflects \method's commitment to evidence-based repair; by grounding the resolution process in a verifiable reproduction script, the system ensures that proposed fixes are validated against the actual failure, significantly reducing the risk of regression or incomplete repair in complex, large-scale repositories.

To ground reproduction in repository semantics, the agent retrieves issue-relevant code and documentation context through the context retrieval engine. This retrieved context is then used to guide the synthesis of reproduction artifacts, including test scripts, driver code, or configuration changes, depending on the project’s build system and testing conventions (e.g., Maven, npm, or custom commands). Reproduction is realized through an iterative write–execute–analyze loop. The agent incrementally modifies the codebase or test suite to introduce reproduction logic, materializes these changes as explicit patches, and deploys them to the execution environment. Within this environment, the agent executes synthesized reproduction commands to empirically validate whether the reported failure can be reliably triggered. Execution results are analyzed by an LLM via simple few-shot prompting that determines whether the bug has been successfully reproduced. If reproduction fails, the agent leverages execution logs and diagnostic feedback from prior attempts to iteratively refine the reproduction procedure. This feedback-driven loop continues until a reproducible failure is established or a predefined iteration budget is reached.

By producing executable reproduction tests and verified failure traces, the Bug Reproduction Agent transforms ambiguous natural-language bug reports into concrete, measurable failure cases. This not only stabilizes the repair objective but also provides a precise baseline against which candidate patches can be evaluated in later stages of the pipeline.

\subsubsection{Patch Generation Agent}

The Patch Generation Agent adapts its orchestration strategy based on the outcome of the Bug Reproduction Agent. We categorize bugs into two types: \textit{verified bugs}, where a reproduction test has successfully triggered the failure, and \textit{unverified bugs}, where a deterministic reproduction could not be established, or the issue pertains to feature requests and refactoring.

For a verified bug, the agent initiates a targeted retrieval process via the Context Retrieval Engine to gather localized context. During the subsequent analysis phase, the LLM is equipped with a web search tool to cross-reference external documentation or similar issues, enabling it to synthesize a comprehensive bug fix plan. Following this plan, the agent iteratively edits the source code to generate a candidate patch. If the generated patch does not pass through the Patch Verification Agent, the Patch Generation Agent re-enters the analysis loop to further refine its fix. This iterative generation process continues until a plausible patch is produced. In contrast, for an unverified bug, the agent follows a modified workflow that bypasses test-grounded reasoning. While the initial context retrieval and planning phases remain consistent, the agent compensates for the lack of test-based validation by executing the generation process multiple times. This results in a diverse set of candidate patches, allowing the system to explore multiple potential fix hypotheses through repeated sampling rather than relying on a single iterative refinement loop.

\subsubsection{Patch Verification Agent}

The Patch Verification Agent evaluates the correctness and robustness of candidate patches generated by the Patch Generation Agent.

For verified bugs, each candidate patch is validated by executing both the reproduction test and a selected set of regression tests. 
Candidate regression tests are first retrieved by the Context Retrieval Engine based on their semantic relevance to the modified code regions. 
An LLM then selects a focused subset of these tests that are most likely to be impacted by the patch and thus informative for validation.
A patch is immediately accepted as the final fix once it passes both the reproduction test and all selected regression tests. 
Otherwise, the patch is discarded, and control is returned to the Patch Generation Agent. This generate–verify loop continues until a valid patch is found or the iteration budget is exhausted.

For unverified bugs as well as feature or refactoring issues, the agent relies solely on selected regression tests for validation. 
All patches that pass regression testing are collected into a verified patch pool, which is subsequently normalized and deduplicated to remove semantically equivalent candidates. 
Normalization canonicalizes patch representations (e.g., formatting and edit ordering), while deduplication is approximated using structural similarity of code edits.
From the remaining candidates, the most suitable patch is selected via an LLM-based aggregation mechanism and majority voting.
This design follows prior work such as TRAE~\cite{gao2025trae} and Agentless~\cite{xia2024agentless}, which employ multi-patch generation and aggregation when executable verification is unavailable.
If no patch passes verification, the same aggregation mechanism is applied over all generated candidates to ensure graceful degradation and the return of a plausible solution.

\section{Experimental Methodology}
\label{sec:evaluation}

\subsection{Research Questions}
\begin{itemize}

\item \textbf{RQ1 (Effectiveness Comparison)}: How effectively does \method resolve issues compared to related code agents?

\item \textbf{RQ2 (Analysis of Context Retrieval)}: What's the quality of \method's context retrieval compared to prior techniques?

\item \textbf{RQ3 (Ablation Study of Core Components)}: 
How does removing key components of \method—namely, \textit{working memory}, \textit{multiple patch selection}, \textit{bug reproduction agent}, and \textit{regression testing}—affect overall issue resolution performance and reliability?

\end{itemize}

\subsection{Experimental Setup}

\method is implemented in Python, and uses GPT-5 (temperature 1.0) for generation and reasoning, and \texttt{codestral-embed-2505} for subquery embedding in working memory. Experiments are conducted on an Ubuntu 24.04 server with 32 CPU cores (Intel Xeon E5-2667 v4), 125\,GiB RAM. 
We utilize Neo4j for Knowledge Graph storage and PostgreSQL for working memory. For context retrieval, AST nodes are chunked at traversal depth 1, while documentation uses 5,000-token chunks (500-token overlap). The working memory module filters context using a 0.85 similarity threshold, retaining the top-5 entries. We generate up to 5 candidate patches per issue, all evaluated in a uniform environment to ensure reproducibility. Both Bug Reproduction and Patch Verification Agents operate within isolated Docker containers to ensure environment consistency and security. 
For the web search tool, the agent integrates Tavily~\cite{tavily} as its web search tool to obtain up-to-date documentation and repository-related knowledge when required.

\subsubsection{Benchmarks}

\begin{table}[t]
\centering
\small
\caption{Statistics of evaluation benchmarks used in our work.}
\label{tab:benchmark_overview}
\setlength{\tabcolsep}{7pt}
\renewcommand{\arraystretch}{1.2}
\begin{tabular}{lccc}
\toprule
\textbf{Benchmark} & \textbf{Tasks} & \textbf{Languages} & \textbf{Task Types} \\
\midrule
\text{SWE-bench Verified} &
500 &
Python &
Bug Fixing \\
\text{SWE-PolyBench Verified} &
382 &
\makecell{Python, Java, \\ JavaScript, TypeScript} &
\makecell{Bug Fixing, Feature \\ Implementation, Refactoring} \\
\bottomrule
\end{tabular}
\end{table}

We evaluate \method through a series of comprehensive experiments designed to assess its efficacy, stability, and generalization capabilities in automated software engineering. As shown in~\autoref{tab:benchmark_overview}, our evaluation centers on two primary benchmarks: SWE-bench Verified~\cite{jimenez2024swebench}, a human-curated subset of 500 instances from the original SWE-bench characterized by high-quality problem statements and reliable unit tests for Python-based issue resolution, and SWE-PolyBench Verified~\cite{rashid2025swepolybench}, a multi-language benchmark of 382 instances that extends evaluation beyond bug fixing to include feature implementation and refactoring across 3 additional languages, such as Java, JavaScript, and TypeScript. Both benchmarks provide only the issue description and the full repository, without explicit fault localization or additional oracle information. Together, these benchmarks enable a robust assessment of the agent’s generality and its ability to adapt to heterogeneous, real-world codebases. Systems are therefore required to reason over repository-level context, reflecting realistic software maintenance settings.

\subsubsection{Methodology}
\paragraph{Effectiveness Comparison (RQ1)}
To evaluate the effectiveness of \method, we evaluate it on two real-world software engineering benchmarks: SWE-bench Verified~\cite{jimenez2024swebench} and SWE-PolyBench Verified~\cite{rashid2025swepolybench}.

We compare \method with several strong baselines reported in prior work and public leaderboards, namely \textit{OpenHands}~\cite{wang2025openhands} \textit{Agentless}~\cite{xia2024agentless}, \textit{live-SWE-agent}~\cite{xia2025live}, and \textit{SWE-agent}~\cite{yang2024sweagent} on SWE-bench Verified, and Amazon Q Developer Agent~\cite{amazonQDeveloper2025}, Aider~\cite{aider2024}, SWE-agent~\cite{yang2024sweagent}, and Agentless~\cite{xia2024agentless} on SWE-Polybench. We do not reimplement any baseline systems; instead, we directly adopt results reported in the corresponding papers or official benchmark leaderboards. We report the percentage of resolved issues as the primary evaluation metric to assess the effectiveness of \method under identical benchmark definitions.

\paragraph{Analysis of Context Retrieval (RQ2)}

To analyze how the context retrieval mechanism of \method differs from existing agents, we conduct a trajectory-based comparative analysis focusing on the quality of retrieved context rather than end-to-end task success.

We randomly select 100 instances from SWE-bench Verified due to limited resources, and we run these instances using GPT-5 on \method, \textit{Agentless}~\cite{xia2024agentless}, \textit{OpenHands}~\cite{wang2025openhands}, and \textit{SWE-agent}~\cite{yang2024sweagent}. We selected these agents because they are open-sourced and well-maintained. For each instance, we use the gold patch provided by the benchmark to construct a gold context. Starting from the files and code regions modified in the gold patch, we engage six professional developers to manually trace the relevant functions and dependency-related code required to understand and implement the fix, yielding a minimal gold context that generates a valid patch. For each agent, we extract all contextual artifacts retrieved throughout its execution trajectory, including files, functions, and code spans accessed via search, ranking, or navigation actions, and aggregate them to form the agent-retrieved context. We then compare the agent-retrieved context against the corresponding gold context to assess context retrieval accuracy and characterize retrieval behavior across agents.

\paragraph{Ablation Study of Core Components (RQ3)}
We consider four core components: \textit{working memory}, \textit{multiple patch selection}, \textit{bug reproduction agent}, and \textit{regression testing}. Starting from the full \method configuration, we create ablated variants by disabling each component independently, while keeping all other components, model backbones, prompts, and execution settings unchanged. This controlled setup allows us to isolate the impact of each component on overall system behavior. Each ablated variant is evaluated under the same benchmark protocols as the full system. We measure issue resolution performance using the percentage of resolved issues. By comparing the ablated variants against the full configuration, we quantify how the absence of individual components affects both effectiveness and reliability in repository-level issue resolution.


\section{Experimental Results}
\subsection{RQ1: Effectiveness Comparison}

\begin{table}[!t]
\centering
\small
\caption{\method Performance on SWE-bench Verified.}
\label{tab:swebench_verified}
\setlength{\tabcolsep}{6pt}
\renewcommand{\arraystretch}{1.05}
\begin{tabular}{p{6.5cm} c c}
\toprule
\textbf{Agent} & 
\textbf{Resolve Rate (\%)} \\
\midrule
SWE-Exp                                          & 42.0 \\
Agentless-1.5 + Claude-3.5 Sonnet (20241022)     & 50.8 \\
mini-SWE-agent + Gemini 2.5 Pro (2025-05-06)     & 53.6 \\
mini-SWE-agent + DeepSeek V3.2 Reasoner          & 60.0 \\
mini-SWE-agent + GPT-5                           & 65.0 \\
SWE-agent + Claude 4 Sonnet                      & 66.6 \\
OpenHands + Claude 4 Sonnet                      & 70.4 \\
Lingxi v1.5 + Kimi K2                            & 71.2 \\
OpenHands + GPT-5                                & 71.8 \\
\hline
\textbf{\method + GPT-5 (Our Work)}              & \textbf{74.4} \\
\bottomrule
\end{tabular}
\end{table}

\begin{table}[!t]
\centering
\small
\caption{\method Performance on SWE-PolyBench Verified.}
\label{tab:swepolybench_verified}
\renewcommand{\arraystretch}{1.05}
\begin{tabular}{p{6cm} c c c c c}
\toprule
\textbf{Agent} &
\textbf{Overall} &
\textbf{Java} &
\textbf{Python} &
\textbf{JavaScript} &
\textbf{TypeScript} \\
\midrule

\multicolumn{6}{c}{\textbf{Part A: Overall and Language-Wise Resolution Rate (\%)}} \\
\midrule
Aider-PB (Mistral-Large)             & 8.4  & 10.1 & 9.7  & 5.0  & 9.0  \\
AgentlessPB (Sonnet 3.5)             & 13.3 & 17.4 & 23.0 & 7.0  & 6.0  \\
Aider-PB (Deepseek R1)               & 13.9 & 8.7  & 18.6 & 14.0 & 12.0 \\
SWE-agent-PB (Sonnet 3.5)            & 14.4 & 18.8 & 22.1 & 5.0  & 12.0 \\
Aider-PB (Sonnet 3.5)                & 16.2 & 20.3 & 20.4 & 11.0 & 14.0 \\
Amazon Q Developer Agent (v20240402) & 28.8 & 37.7 & 35.4 & 20.0 & 24.0 \\
\midrule
\textbf{\method+ GPT-5 (Our Work)}     & \textbf{33.8} & \textbf{33.3} & \textbf{36.3} & \textbf{30.0} & \textbf{35.0} \\
\midrule
\multicolumn{6}{c}{\textbf{Part B: Task-Wise Resolve Rate of \method (\%)}} \\
\midrule
Bug Fixing                & \multicolumn{5}{c}{37.5} \\
Feature Implementation    & \multicolumn{5}{c}{18.6} \\
Refactoring               & \multicolumn{5}{c}{30.8} \\
\bottomrule
\end{tabular}
\vspace{-1em}
\end{table}

As shown in~\autoref{tab:swebench_verified} and \autoref{tab:swepolybench_verified}, we compare \method against a diverse set of representative agents on two widely used benchmarks: SWE-bench Verified~\cite{jimenez2024swebench} and SWE-PolyBench Verified~\cite{rashid2025swepolybench}. SWE-bench Verified focuses on single-language, execution-verified bug-fixing tasks, while SWE-PolyBench evaluates robustness in a more challenging multi-language and multi-task setting. Together, these benchmarks provide a comprehensive view of effectiveness and generalization.

On SWE-bench Verified, \method achieves a resolution rate of 74.4\%, placing it among the top-performing systems. It outperforms a wide range of strong baselines, including fully autonomous agents such as OpenHands~\cite{wang2025openhands} with GPT-5 (71.8\%) and SWE-agent~\cite{yang2024sweagent} with Claude 4 Sonnet (66.6\%), as well as recent competitive systems such as Lingxi v1.5~\cite{yang2025lingxi} + Kimi K2 (71.2\%). \method also consistently surpasses lighter-weight or search-based approaches, including Agentless-1.5~\cite{xia2024agentless} (50.8\%) and multiple mini-SWE-agent variants using Gemini 2.5 Pro (53.6\%), DeepSeek V3.2 Reasoner (60.0\%), and GPT-5 (65.0\%). These results indicate that the performance gains of \method cannot be attributed solely to model choice, but rather stem from its system-level design. 

We further compare the sets of unique resolved instances across \method, OpenHands, and mini-SWE-agent using the same GPT-5 model. \method resolves the largest number of unique instances (29), followed by OpenHands (16) and mini-SWE-agent (4). The larger unique region associated with \method indicates that its performance gains stem not only from common cases, but also from resolving harder issues. We attribute this advantage to its repository-level Knowledge Graph and Memory Enhanced Context Retrieval Engine, which supports more reliable long-horizon reasoning.

As shown in~\autoref{tab:swepolybench_verified}, \method attains an overall resolution rate of 33.8\% on SWE-PolyBench Verified, ranking first on this more challenging benchmark. It outperforms other strong baselines, including Amazon Q Developer Agent~\cite{amazonQDeveloper2025} (28.8\%), and more than doubles the performance of widely used tools such as Aider-PB~\cite{aider2024} with Sonnet~3.5 (16.2\%). In particular, \method substantially surpasses SWE-agent-PB~\cite{yang2024sweagent} (14.4\%) and AgentlessPB~\cite{xia2024agentless} (13.3\%), demonstrating stronger robustness in multi-language and multi-task settings. Task-wise analysis further shows that \method performs best on bug-fixing tasks (37.5\%), while also achieving strong results on refactoring (30.8\%) and maintaining competitive performance on feature implementation tasks (18.6\%). To the best of our knowledge, \method is the first work to report task-wise resolution rates on SWE-Polybench.

Overall, these results show that \method goes beyond patch generation, exhibiting strong capabilities in long-horizon issue resolution and repository-level understanding. By integrating repository-level Knowledge Graph with a Memory Enhanced Context Retrieval Engine, \method resolves a broader range of issues, including many unsolved by prior agent frameworks. This marks a clear shift from patch-centric repair toward holistic, program-level reasoning, bringing automated program repair closer to practical, real-world applicability.

\begin{summary}
\textbf{Answer to RQ1:}
\method achieves the highest resolution rates on both benchmarks and resolves the most unique instances.
This advantage extends to harder and multi-language issues, indicating stronger generalization than prior agents.
These results highlight the effectiveness of repository-level knowledge and memory-enhanced retrieval.
\end{summary}

\subsection{RQ2: Analysis of Context Retrieval}

\begin{table*}[t]
\centering
\small
\setlength{\tabcolsep}{7pt}
\caption{
Average context hit rate comparison at file, class/function, and span levels.
}
\begin{tabular}{lccc}
\toprule
\textbf{Agent} &
\textbf{File Hit Rate$\uparrow$} &
\textbf{Class/Function Hit Rate$\uparrow$} &
\textbf{Span Hit Rate$\uparrow$} \\
\midrule
Agentless  & 0.802 & 0.395 & 0.054 \\
OpenHands  & 0.876 & 0.579 & 0.541 \\
SWE-Agent  & 0.632 & 0.524 & 0.482 \\
\hline
\method & \textbf{0.915} & \textbf{0.850} & \textbf{0.807}  \\
\bottomrule
\end{tabular}
\vspace{1pt}

\label{tab:contextbench_main}
\end{table*}

We conduct a qualitative analysis of context retrieval quality for representative top-performing agents on SWE-bench~\cite{jimenez2024swebench}, examining their ability to retrieve relevant context at multiple level. As shown in~\autoref{tab:contextbench_main}, \method consistently outperforms all baselines across file-, symbol-, and span-level hit rates, indicating a substantially stronger alignment between retrieved context and the gold patch evidence.

At the file level, \method achieves a hit rate of 0.915, surpassing OpenHands (0.876) and Agentless (0.802), suggesting that it more reliably identifies the set of files relevant to the underlying issue. This advantage becomes more pronounced at finer granularities. At the symbol level, \method attains a hit rate of 0.850, representing a large margin over all baselines, which remain below 0.58. This result indicates that \method is markedly more effective at recovering semantically relevant functions, classes, and variables, rather than relying primarily on coarse file-level context. The gap widens further at the span level, where \method reaches a hit rate of 0.807, substantially outperforming OpenHands (0.541), SWE-Agent (0.482), and especially Agentless (0.054). This demonstrates \method’s superior ability to localize fine-grained code regions that directly correspond to the modifications made in the gold patches. Such fine-grained retrieval is critical for downstream patch generation, as precise localization of relevant code spans provides stronger and less ambiguous evidence for reasoning about bug causes and fixes.

Overall, these results highlight that \method’s retrieval mechanism is not only broader in coverage but also more structurally precise across multiple levels of granularity. By effectively bridging file-level discovery and span-level localization, \method enables a more comprehensive and actionable understanding of the program context, which is essential for reliable long-horizon issue resolution compared to prior agent-based approaches.

\begin{summary}
\textbf{Answer to RQ2:}
\method consistently retrieves more accurate context across file, symbol, and span levels than prior agents.
Its advantage is especially pronounced at fine-grained symbol and span levels, indicating superior localization of patch-relevant code.
This precise retrieval provides stronger evidence for downstream reasoning and patch generation.
\end{summary}

\subsection{RQ3: Ablation Study of Core Components}

\begin{figure}[t]
    \centering
    \includegraphics[width=0.48\linewidth]{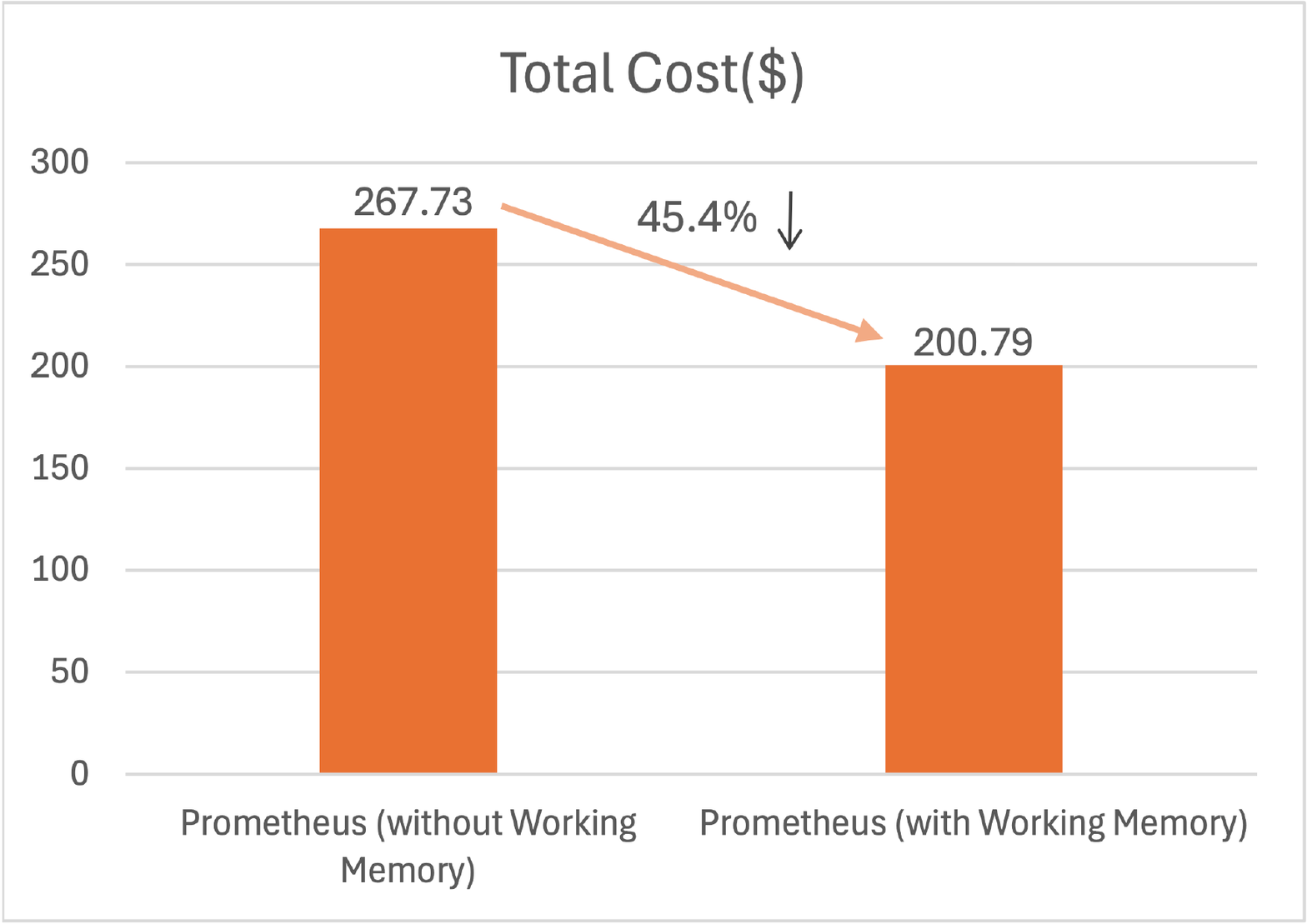}
    \hfill
    \includegraphics[width=0.48\linewidth]{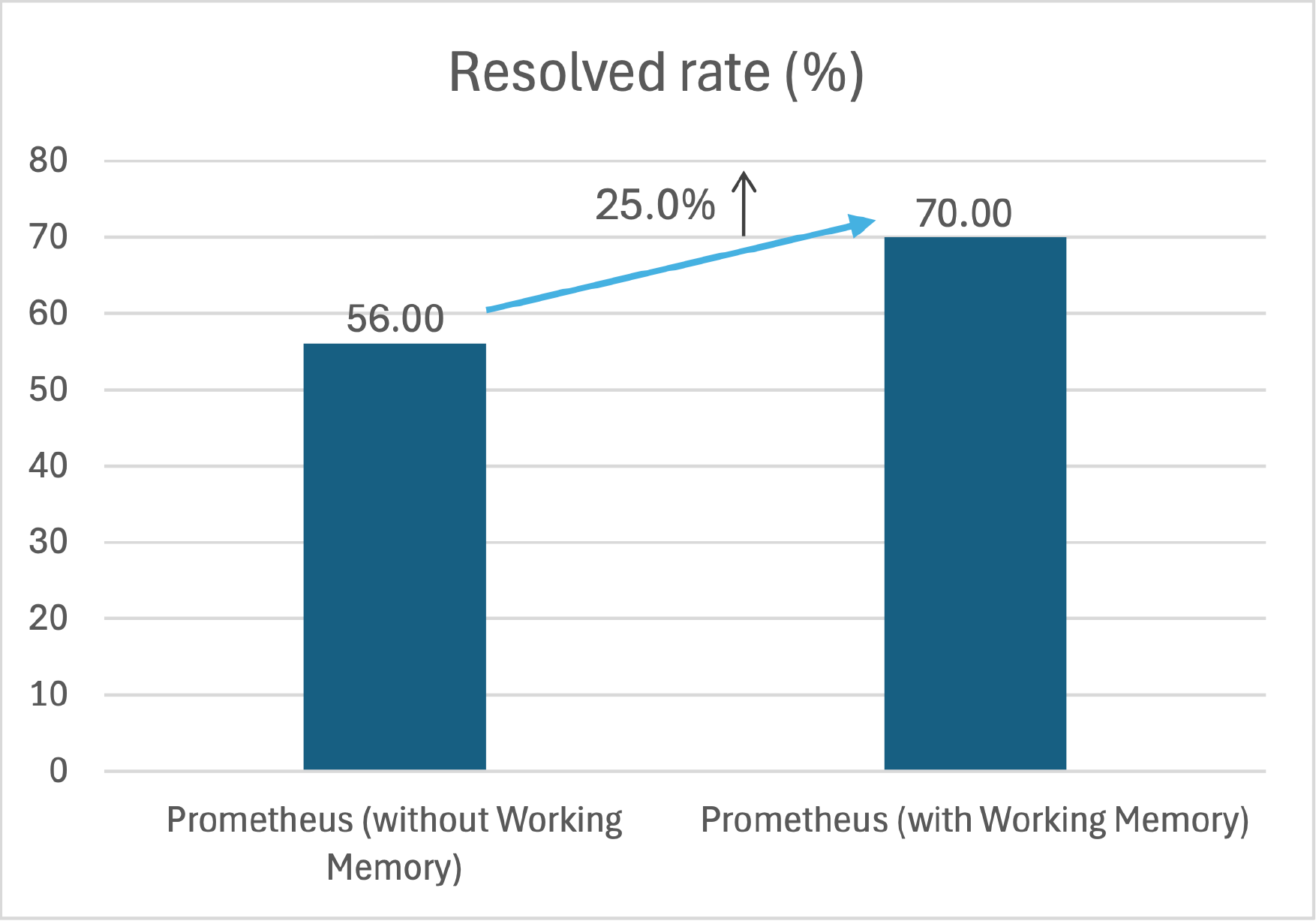}
    \caption{Ablation study on working memory.}
    \label{fig:prometheus_without_memory}
\end{figure}

\subsubsection{Ablation Study on Working Memory}

To evaluate the efficiency of the working memory mechanism, we have done an ablation study by removing it.
Our results show that incorporating working memory substantially improves both efficiency and performance. Due to limited resources, we randomly select 50 instances from SWE-bench Verified~\cite{jimenez2024swebench} and run them in both configurations. As summarized in~\autoref{fig:prometheus_without_memory}, \method with working memory achieves a resolution rate of 70.00\% while incurring a total inference cost of \$200.79 across 50 issues. In contrast, the variant without working memory resolves only 56.00\% of issues and incurs a significantly higher cost of \$367.73. This demonstrates that working memory yields significant resource savings, reducing inference cost by approximately 45.4\%~$\downarrow$, while simultaneously improving the issue resolution rate by 25.0\%~$\uparrow$. This demonstrates that working memory effectively reduces redundant context retrieval and excessive token consumption, while stabilizing long-horizon reasoning by preserving query-relevant context across iterative repair steps. Without working memory, repeated retrieval and short-horizon context not only increases inference cost but also degrades decision quality. 

\parabf{Case Study: Impact of Working Memory on Retrieval Efficiency}
We present a representative case from SWE-bench Verified to illustrate how working memory improves retrieval efficiency and stabilizes long-horizon reasoning.
The issue requests setting a default value of \texttt{FILE\_UPLOAD} \texttt{\_PERMISSIONS} to \texttt{0o644}, motivated by inconsistent file permissions caused by different upload handlers and temporary file backends.

In the configuration without working memory, the agent repeatedly retrieved the same core files across multiple iterations, including
\texttt{django/core/files/storage.py},
\texttt{django/core/files} \texttt{/uploadhandler.py}, and
\texttt{django/core/files/uploadedfile.py}.
Because previously retrieved context was not preserved, each reasoning step re-triggered similar repository-level searches and re-encoding of identical code regions.
As the agent attempted to reconcile interactions between upload handlers, temporary file creation, and final storage semantics, this repeated retrieval caused substantial token overhead and context redundancy.
Moreover, the lack of persistent memory led to fragmented reasoning about the permission propagation path, resulting in unstable patch exploration and significantly increased inference cost.

When working memory is enabled, the same files are retrieved once and stored as query-relevant context.
Subsequent reasoning steps directly reuse these memory entries rather than re-fetching identical code from the repository.
This allows the agent to maintain a coherent view of how upload handlers, temporary files, and storage backends jointly affect file permissions.
In addition, related contextual information—such as where permissions are ultimately applied during file saving—is retained across iterations, enabling more focused retrieval of only genuinely new context.
As a result, the agent converges more quickly to a consistent solution while substantially reducing redundant token consumption.

\subsubsection{Ablation Study on Multiple Patch Selection}

\begin{table}[t]
\centering
\small
\caption{Ablation study of core components of \method.}
\label{tab:ablation_components}
\setlength{\tabcolsep}{8pt}
\renewcommand{\arraystretch}{1.1}
\begin{tabular}{l c c}
\toprule
\textbf{Removed Component} & \textbf{Resolve Rate (\%)}  & \textbf{Performance Drop}\\
\midrule
No Component Removed     & 74.40 & -\\ 
Regression Testing       & 70.20 & 5.6\%~$\downarrow$ \\
Multiple Patch Selection & 68.80 & 7.0\%~$\downarrow$ \\
Bug Reproduction         & 64.00 & 14.0\%~$\downarrow$\\
\bottomrule
\end{tabular}
\end{table}

To evaluate the contribution of \textit{multiple patch selection}, we conduct an ablation experiment by disabling this component. We evaluate this variant on all 500 instances from SWE-bench Verified~\cite{jimenez2024swebench}. As a result, the overall resolution rate score drops from 74.4 for the full system to 69.2 without multiple patch selection, corresponding to a degradation of 7.0\%~$\downarrow$. The results highlight the importance of multi-patch selection. This approach allows the system to evaluate diverse candidates, preventing it from settling on an incorrect solution. These results suggest that aggregating and selecting among multiple candidate patches substantially enhances robustness and reliability in agentic issue resolution.

\subsubsection{Ablation Study on Bug Reproduction}

To assess the impact of the \textit{Bug Reproduction Agent}, we perform an ablation study by disabling the reproduction test. We evaluate this variant on the full SWE-bench Verified benchmark. The resolution rate score decreases from 74.4 with the full system to 64.0 without bug reproduction, representing a substantial drop of 14.0\%~$\downarrow$. This substantial drop highlights the importance of explicitly reproducing reported bugs, as reproduction tests provide a strong execution-based signal for verifying whether a candidate patch truly addresses the root cause of the issue. Without this signal, the system is more prone to accepting patches that pass partial validation but fail to resolve the original bug, leading to reduced effectiveness and reliability.

\subsubsection{Ablation Study on Regression Testing}

To examine the role of \textit{regression testing component}, we conduct an ablation study by removing this component from \method and evaluating candidate patches without executing additional regression tests beyond bug reproduction. We evaluate this variant on the full SWE-bench Verified benchmark. The resolution rate score decreases from 74.4\ with the full system to 70.2 without regression testing, corresponding to a reduction of 5.6\%~$\downarrow$. This result demonstrates that regression tests provide an important complementary verification signal, helping to detect patches that fix the target bug but inadvertently break existing functionality. Removing regression testing increases the risk of accepting overfitted or brittle patches, leading to reduced overall reliability despite a smaller performance drop compared to removing bug reproduction.

\begin{summary}
\textbf{Answer to RQ3:}
Removing any core component leads to a clear performance drop, with bug reproduction having the largest impact.
Multiple patch selection and regression testing further improve robustness by preventing premature or overfitted patches.
Working memory significantly reduces cost while improving resolution, highlighting its role in long-horizon reasoning.
\end{summary}

\section{Related Work}
\label{related_work}

\subsection{Agentic Issue Resolution}

The field of software maintenance is transitioning from traditional Automated Program Repair (APR) toward comprehensive agentic issue resolution. Historically, APR focused on generating patches for localized bugs, relying on predefined patterns \cite{liu2019tbar} or heuristics \cite{le2011genprog,YE2021110825} validated against functional-level test oracles \cite{iter}. While effective for well-scoped bugs, these methods struggle with the complexity of repository-level issues. Recent advancements in LLMs have catalyzed a shift toward autonomous code agents capable of long-horizon reasoning across multiple components \cite{jimenez2024swebench, xia2023automated}. Existing coding agents can be categorized into two primary paradigms based on their orchestration: \textit{fully-autonomous agents} and \textit{workflow-based agents}.

\paragraph{Fully-autonomous Agent} This paradigm utilizes LLMs to dynamically determine execution workflows, allowing for flexible tool invocation based on the evolving context of a task. SWE-Agent~\cite{yang2024sweagent} is a pioneer in this category, introducing the Agent-Computer Interface (ACI) to facilitate more effective interaction between the agent and the software environment. OpenHands~\cite{wang2025openhands} builds upon the foundation of an ACI similar to SWE-Agent. ClaudeCode \cite{anthropic2025claudecode} further pushes the boundaries of agency by integrating deep reasoning with a CLI-based execution loop for real-time, autonomous repository maintenance. Codex \cite{openai2025codex} represents a significant milestone, demonstrating sophisticated autonomous problem-solving capabilities across large-scale codebases. Expanding this autonomy further, Live-SWE-Agent \cite{xia2025live} introduces a self-evolving paradigm that allows the agent to autonomously and continuously refine its own scaffold implementation on-the-fly during runtime. By evolving from basic tools to complex scaffolds without offline training, it has achieved state-of-the-art performance on benchmarks such as SWE-bench Verified~\cite{jimenez2024swebench}. While highly flexible, these fully autonomous agent systems often struggle with non-deterministic execution paths and significant computational overhead.

\paragraph{Workflow-based Agent} Workflow-based (or pipeline-based) agents follow a predefined, human-designed workflow to ensure stability, efficiency, and reproducibility. Agentless \cite{xia2024agentless} is a representative method that intentionally avoids complex agentic loops, instead following a strict "localization-generation-validation" sequence. AutoCodeRover \cite{zhang2024autocoderover} and SpecRover \cite{ruan2024specrover} enhance this process by combining spectrum-based fault localization with LLM-guided code search within a structured pipeline. TRAE \cite{gao2025trae} and Lingxi \cite{yang2025lingxi} also adopt this paradigm, utilizing more sophisticated but fixed orchestration layers to manage large-scale repository edits. By constraining the agent's action space to a structured execution graph, these workflow-based systems mitigate the risks of "hallucinated" workflows and significantly reduce the search space for software patches.

Despite their success, existing coding agents—from \textit{fully-autonomous} loops to \textit{workflow-based} pipelines—frequently encounter a "retrieval bottleneck" in large codebases due to a lack of structured memory and iterative refinement. Unlike prior works that rely on one-off tool invocations, \method utilizes a special context retrieval mechanism integrated with a multi-vector working memory. By transitioning from static search to iterative, memory-augmented traversal, our approach ensures precise evidence collection across complex repository structures.

\subsection{Repository-level Context Retrieval}

Effective issue resolution in real-world software repositories often requires coherent, repository-level context, as bugs frequently span multiple files, cross-file dependencies, and multi-layer abstractions. To address this challenge, recent work has explored repository-level context retrieval as a core primitive for downstream tasks such as code completion, understanding, and repair. Focusing on efficiency, REPOFUSE~\cite{liang2024repofuserepositorylevelcodecompletion} retrieves analogy- and rationale-based context while optimizing for latency and token usage. Similarly, RepoCoder~\cite{zhang-etal-2023-repocoder} leverages iterative retrieval to capture semantically related code chunks across the repository. Beyond raw code, DocPrompting~\cite{DocPrompting} retrieves documentation-relevant snippets via embedding-based techniques, while LTFix~\cite{LTFix} targets memory errors in C codebases by retrieving typestate-guided usage traces.

To capture the structural complexity of codebases, recent research has turned to Knowledge Graphs. RepoGraph~\cite{ouyang2024repograph} models code entities and their interactions as graphs, while KGCompass~\cite{yang2025enhancing} further links code with repository artifacts such as issues and pull requests to enable path-based reasoning. In contrast, Code Graph Model~\cite{tao2025code} embeds repository-level code graphs directly into LLMs to support joint semantic and structural reasoning. By formalizing repositories as structured semantic networks, knowledge graph-based methods provide a coherent architectural map that improves contextual integrity for downstream tasks.

\method features a language-agnostic graph model for repository-scale retrieval across diverse tasks. While agents such as OpenHands~\cite{wang2025openhands} are efficient executors, they often lack deep contextual reasoning for long-horizon issues. In contrast, \method incorporates a \emph{working memory} mechanism to persistently maintain and selectively update salient context during issue resolution, reducing redundant retrieval and repeated reasoning while improving logical consistency and computational efficiency.

\subsection{Memory for Agents}
Recent agent-based systems increasingly incorporate explicit memory mechanisms to support long-horizon reasoning, contextual consistency, and continual improvement, moving beyond treating memory as a transient by-product of the context window.

MemGPT~\cite{packer2023memgpt} introduces an OS-inspired memory hierarchy that separates a limited working context from persistent external storage, enabling agents to retain long-term factual information through explicit read and write operations. Voyager~\cite{wang2023voyager} accumulates reusable skills distilled from successful task trajectories, allowing agents to progressively expand procedural capabilities without retraining. While these memory mechanisms are primarily studied in agents with other purposes or tasks, similar ideas have recently been adapted to software issue resolution, where agents must reason over large codebases and maintain evolving repair context across multiple steps. EXPEREPAIR~\cite{mu2025experepair} and SWE-Exp~\cite{chen2025swe} explicitly model cross-issue repair experience by distilling reusable repair demonstrations and abstracted problem-solving insights from prior issue-resolution trajectories, which are dynamically retrieved to guide future repository-level fixes. RepairAgent~\cite{bouzenia2024repairagent} maintains a dynamically updated interaction state within a single repair episode, preserving gathered code context, hypotheses, and tool outputs to support iterative decision making, but does not persist experience across issues.

In contrast to these approaches, \method adopts a system-oriented working memory design tailored to software repositories. Rather than maintaining issue-local interaction traces or retrieving past trajectories, its working memory explicitly organizes and updates query-relevant repository artifacts (e.g., files, functions, and tests) and evolving repair states across iterative steps, enabling semantically grounded context management for long-horizon software engineering tasks.

\section{Threats to Validity}

\paragraph{Internal Validity.}
A potential internal threat lies in whether the observed improvements truly result from the proposed memory mechanism rather than random factors or implementation variance.
Due to the high computational and API costs of running LLM-based agents, we conduct each experiment once under strictly controlled settings, using fixed random seeds, identical prompts, and consistent tool configurations across all baselines.
Although this limitation prevents repeated trials, the performance differences between \method and prior agents are substantial and consistent across tasks, suggesting that the observed trends are robust.

\paragraph{External Validity.}
External validity threats concern the generalizability of our findings to other repositories, programming languages, and LLM backbones.
To mitigate this threat, we evaluate \method on two representative issue resolution benchmarks, SWE-bench Verified and SWE-PolyBench Verified, where the latter covers multiple programming languages and diverse repository structures.
Although these benchmarks may not fully capture the scale and complexity of real-world industrial systems, the consistent performance of \method across both datasets and multiple programming languages suggests that the findings generalize well, thereby alleviating this threat.

\paragraph{Construct Validity.}
Construct validity concerns whether our evaluation metrics accurately reflect the capability of long-horizon codebase navigation.
We measure issue resolution accuracy using standardized success criteria defined in SWE-bench and SWE-PolyBench, and further assess context retrieval precision against human-annotated gold contexts to capture the quality of contextual reasoning.
Although these metrics may not fully represent all qualitative aspects of real-world software development, their consistency across both task-level and retrieval-level evaluations provides mutual validation, mitigating this threat.

\section{Conclusion}

We presented \method, a multi-agent system that shifts the focus of automated software issue resolution from local patch generation to long-horizon, repository-level software engineering. By integrating repository-level knowledge graphs with a memory-enhanced retrieval engine, \method enables deep program understanding and iterative reasoning across complex codebases. Through the coordination of specialized agents for retrieval, generation, and verification, the system effectively unifies structural dependencies, historical context, and execution feedback. Our evaluations on SWE-bench Verified and SWE-PolyBench Verified demonstrate that \method significantly outperforms existing baselines, particularly in challenging multi-language and multi-task scenarios. These results underscore that robust automated maintenance requires principled mechanisms for long-horizon reasoning and repository-level abstraction beyond simple code generation. \method represents a critical step toward practical, reliable software agents that align with real-world development workflows.

\section{Data Availability}
All the experimental data and code used in this paper are available at  \texttt{\url{ https://github.com/EuniAI/Prometheus}}.

\bibliography{custom}
\bibliographystyle{ACM-Reference-Format}

\newpage
\appendix
\onecolumn
\end{document}